\documentclass[12pt]{article}

\newcommand{\pslash}{\not \! p}

\usepackage{amsmath,amssymb,amscd,amsbsy,amsgen,amsopn,amstext,
amsxtra}

\usepackage[dvips]{graphicx}

\begin{document}

\begin{flushright}
\end{flushright}

\vskip 0.5 truecm

\begin{center}
{\Large{\bf Lorentz invariant CPT violation: \\Particle and antiparticle mass splitting}}
\end{center}
\vskip .5 truecm
\begin{center}
{\bf { Masud Chaichian{$^*$}, Kazuo Fujikawa $^\dagger$ and
Anca Tureanu$^*$}}
\end{center}

\begin{center}
\vspace*{0.4cm} {\it { $^*$Department of Physics, University of Helsinki, P.O.Box 64, FIN-00014 Helsinki,
Finland\\
$^\dagger$ Institute of Quantum Science, College of Science and Technology,\\ Nihon University, Chiyoda-ku, Tokyo 101-8308, Japan}}
\end{center}

\makeatletter
\@addtoreset{equation}{section}
\def\theequation{\thesection.\arabic{equation}}
\makeatother

\begin{abstract}
The interpretation of neutrino oscillation data has led to the question whether, in
principle, an antiparticle like antineutrino  can have a different mass than its
particle. In the framework of a Lorentz invariant CPT violation, which is based on
the nonlocal interaction vertex and characterized by the infrared
divergent form factor, we present an explicit Lagrangian model for the
fermion and antifermion mass splitting.
\end{abstract}

\section{Introduction}
It is important  to study the possible violation of CPT
symmetry~\cite{pauli}, in particular, in the framework of Lorentz
invariant theory.
A Lorentz invariant CPT violation, which may be termed as the long
distance CPT violation in contrast to the familiar short distance CPT
violation~\cite{ellis}, has been recently proposed  in \cite{chaichian}.

The interest in CPT violation and its possible implication on Lorentz invariance
breaking  has been recently revived due to the neutrino oscillation experiments, whose
theoretical interpretation
is favoured if the muon antineutrino mass were different than muon neutrino mass \cite{murayama,altarelli,adamson}.

The scheme proposed in \cite{chaichian} is based on the nonlocal interaction vertex and
characterized by the infrared divergent form factor.
To be definite, the idea is illustrated by  the Yukawa-type Lagrangian
\begin{eqnarray}\label{(1.1)}
{\cal L}&=&\bar{\psi}(x)[i\gamma^{\mu}\partial_{\mu}-M]\psi(x) +
\frac{1}{2}\partial_{\mu}\phi(x)\partial^{\mu}\phi(x)-\frac{1}{2}m^{2}\phi^{2}(x)\nonumber\\
&&+ g\bar{\psi}(x)\psi(x)\phi(x)- V(\phi)\\
&&+ g_{1}\bar{\psi}(x)\psi(x)\int
d^{4}y\theta(x^{0}-y^{0})\delta((x-y)^{2}-l^{2})\phi(y).\nonumber
\end{eqnarray}
This Lagrangian is formally Hermitian and the term with a small real
$g_{1}$ and the step function $\theta(x^{0}-y^{0})$ stands for the CPT
and T violating interaction; $l$ is a real constant parameter.

We first note that the present way to introduce CPT violation is based
on the extra form factor in momentum space as
\begin{eqnarray}\label{(1.2)}
&&g_{1}\int d^{4}x\bar{\psi}(x)\psi(x)\int
d^{4}y\theta(x^{0}-y^{0})\delta((x-y)^{2}-l^{2})\phi(y)\nonumber\\
&&=g_{1}\int dp_{1}dp_{2}dq\int
d^{4}x\bar{\psi}(p_{1})e^{-ip_{1}x}\psi(p_{2})e^{-ip_{2}x}\\
&&\times\int
d^{4}y\theta(x^{0}-y^{0})\delta((x-y)^{2}-l^{2})\phi(q)e^{-iqy}\nonumber\\
&&=g_{1}\int dp_{1}dp_{2}dq
(2\pi)^{4}\delta^{4}(p_{1}+p_{2}+q)\bar{\psi}(p_{1})\psi(p_{2})f(q)\phi(q),\nonumber
\end{eqnarray}
where we defined $f(q)\equiv \int
d^{4}z\theta(z^{0})\delta(z^{2}-l^{2})e^{iqz}$,
namely, CPT violation is realized by an insertion of the form factor
$f(q)$ to the $\phi-\psi\bar{\psi}$ coupling in momentum space. The
ordinary local field theory is characterized by $\delta(z)$ and
$f(q)=1$.  The above form factor is infrared divergent, and it is
quadratically divergent in the present example. This infrared divergence
arises from the fact that we cannot divide Minkowski space into
(time-like) domains with finite
{\em 4-dimensional} volumes in a Lorentz invariant manner. The Minkowski
space is hyperbolic rather than elliptic. CPT symmetry is related to the
fundamental structure of Minkowski space, and thus it is gratifying that
its possible breaking is also related to the basic property of Minkowski
space.

For the later use, it is convenient to define the form factors
\begin{eqnarray}\label{(1.3)}
&&f_{\pm}(p)=\int d^{4}z_{1}
e^{\pm ipz_{1}}\theta(z_{1}^{0})\delta((z_{1})^{2}-l^{2}),
\end{eqnarray}
which are  inequivalent for the time-like $p$ due to the factor
$\theta(z_{1}^{0})$.
For the time-like momentum  $p$, one may choose a suitable Lorentz frame
such that $\vec{p}=0$ and
\begin{eqnarray}\label{(1.4)}
f_{\pm}(p^{0})
&=&2\pi \int_{0}^{\infty}dz\frac{z^{2}e^{\pm
ip^{0}\sqrt{z^{2}+l^{2}}}}{\sqrt{z^{2}+l^{2}}},
\end{eqnarray}
and for the space-like momentum  $p$ one may choose a suitable Lorentz
frame such that $p^{0}=0$ and
\begin{eqnarray}
f_{\pm}(\vec{p})\label{(1.5)}
&=&\frac{2\pi}{|p|^{2}}\int_{0}^{\infty} dz\, z \frac{\sin
z}{\sqrt{z^{2}+(|p|l)^{2}}},
\end{eqnarray}
which is analogous to the Fourier transform of the Coulomb potential and
real. The expression  $f_{\pm}(p)$ is mathematically related to the
formula of the two-point Wightman function (for a free scalar field),
which suggests that
$f_{\pm}(p)$ is mathematically well-defined for $p\neq 0$ at least in
the sense of distribution.

The Lagrangian in \eqref{(1.1)} may be quantized by the path integral by
integrating the formal  equations of motion by means of Schwinger's
action principle~\cite{fujikawa}, whose basis is analogous to that of
the  Yang--Feldman formulation~\cite{yang}. We thus have the generating
functional
$\langle 0,+\infty|0,-\infty\rangle_{J}$
with the source term ${\cal
L}_{J}=\bar{\psi}(x)\eta(x)+\bar{\eta}(x)\psi(x)+\phi(x)J(x)$,
and one may  generate Green's functions in a power series expansion of
perturbation as
\begin{eqnarray}\label{(1.6)}
(i)^{n}\langle T^{\star}\phi(x_{1})...\phi(x_{N})\int d^{4}y_{1}{\cal
L}_{I}(y_{1})
....\int d^{4}y_{n}{\cal L}_{I}(y_{n})\rangle,
\end{eqnarray}
where we consider only $N$ scalar particles as external fields, for
simplicity. We use the covariant $T^{\star}$-product which is essential
to make the path integral on the basis of Schwinger's action principle
consistent~\cite{fujikawa}.

On the basis of this quantization, it is confirmed~\cite{CFT} that the
time-reversal non-invariance in the square of the  probability
amplitudes for the
processes $\phi \rightarrow \bar{\psi}\psi$ and its time reversed
formation process,
\begin{eqnarray}\label{(1.7)}
|A(\phi \rightarrow \bar{\psi}\psi)|^{2} \neq |A(\bar{\psi}\psi
\rightarrow \phi)|^{2},
\end{eqnarray}
is realized after averaging over spin directions for the processes $\phi
\rightarrow \bar{\psi}\psi$ and $\bar{\psi}\psi \rightarrow \phi$ as a
result of the interference of two phases $\theta_{i}$ and
$\pm\theta_{CPT}$.  Here $\theta_{i}$ is the dynamical phase of the
Yukawa theory generated by one-loop corrections and $\pm\theta_{CPT}$ is
the phase generated by our CPT- and T-violating interaction. This shows
that the T-violation in \eqref{(1.1)} is genuine. It is convenient to choose
the masses such that $3M> m >2M$, which makes the above decay mode
the only allowed decay mode.

\section{Lagrangian model of fermion mass splitting}

\subsection{Lagrangian formalism}

In the present nonlocal formulation, we have a new possibility which is
absent in a smooth nonlocal extension of the CPT-even local field
theory. The term $i\mu\bar{\psi}(x)\psi(y)$ (to be precise,
$i\mu\bar{\psi}(x)\psi(x)$)
with a real $\mu$ does not appear in the local Lagrangian since it is canceled by its
Hermitian conjugate. Also this term
is CPT-odd. But in the present nonlocal theory one can consider the
Hermitian combination
\begin{eqnarray}\label{(2.1)}
\int d^{4}xd^{4}y[\theta(x^{0}-y^{0})-\theta(y^{0}-x^{0})]\nonumber\\
\times\delta((x-y)^{2}-l^{2})[i\mu\bar{\psi}(x)\psi(y)],
\end{eqnarray}
which is non-vanishing. Under CPT, we have
$i\mu\bar{\psi}(x)\psi(y)\rightarrow-i\mu\bar{\psi}(-y)\psi(-x)$. By
performing the change of integration variables $-x\rightarrow y$ and $-y
\rightarrow x$, this combination  is confirmed to be CPT=$-1$.
In fact, we have the following transformation property of the operator part
\begin{eqnarray}\label{(2.2)}
&& {\rm C}:\ i\mu\bar{\psi}(x)\psi(y)\rightarrow i\mu\bar{\psi}(y)\psi(x),\\
&& {\rm P}:\ i\mu\bar{\psi}(x^{0},\vec{x})\psi(y^{0},\vec{y})\rightarrow
i\mu\bar{\psi}(x^{0},-\vec{x})\psi(y^{0},-\vec{y}),\nonumber\\
&& {\rm T}:\ i\mu\bar{\psi}(x^{0},\vec{x})\psi(y^{0},\vec{y})\rightarrow
-i\mu\bar{\psi}(-x^{0},\vec{x})\psi(-y^{0},\vec{y}),\nonumber
\end{eqnarray}
and thus the overall transformation property is
C=$-1$, P=$1$,  T=$1$. Namely, C=CP=CPT=$-1$.

It is thus interesting to examine a new action
\begin{eqnarray}\label{(2.3)}
S&=&\int d^{4}x\{\bar{\psi}(x)i\gamma^{\mu}\partial_{\mu}\psi(x)
 - m\bar{\psi}(x)\psi(x)\\
 && -\int
d^{4}y[\theta(x^{0}-y^{0})-\theta(y^{0}-x^{0})]\delta((x-y)^{2}-l^{2})\nonumber\\
 &&\times[i\mu\bar{\psi}(x)\psi(y)]\},\nonumber
\end{eqnarray}
which is Lorentz invariant and Hermitian. For the real parameter $\mu$,
the third term has C=CP=CPT=$-1$ and no symmetry to ensure the equality
of particle and antiparticle masses.

The Dirac equation is replaced by
\begin{eqnarray}\label{(2.4)}
&&i\gamma^{\mu}\partial_{\mu}\psi(x)=m\psi(x)\\
&&+i\mu\int
d^{4}y[\theta(x^{0}-y^{0})-\theta(y^{0}-x^{0})]\delta((x-y)^{2}-l^{2})\psi(y).\nonumber
\end{eqnarray}
By inserting an ansatz for the possible solution
\begin{eqnarray}\label{(2.5)}
\psi(x)=e^{-ipx}U(p),
\end{eqnarray}
we have
\begin{eqnarray}\label{(2.6)}
\pslash U(p)&=&mU(p)\nonumber\\
&+&i\mu\int d^{4}y[\theta(x^{0}-y^{0})-\theta(y^{0}-x^{0})]\nonumber\\
&\times&\delta((x-y)^{2}-l^{2})e^{-ip(y-x)}U(p)\nonumber\\
&=&mU(p)
+i\mu[f_{+}(p)-f_{-}(p)]U(p),
\end{eqnarray}
where $f_{\pm}(p)$ is the Lorentz invariant form factor defined in
\eqref{(1.3)}.

The (off-shell) propagator is defined by
\begin{eqnarray}\label{(2.7)}
\int d^{4}x e^{ip(x-y)}\langle
T^{\star}\psi(x)\bar{\psi}(y)\rangle\nonumber\\
=\frac{i}{\pslash-m +i\epsilon-i\mu[f_{+}(p)-f_{-}(p)]},
\end{eqnarray}
which is manifestly Lorentz covariant. Note that we use the
$T^{\star}$-product for the path integral in accord with Schwinger's
action principle, which is based on the equation of motion \eqref{(2.4)}
with a source term added:
\begin{eqnarray}\label{(2.8)}
\langle 0,+\infty|0,-\infty\rangle_{J}=\int{\cal D}\bar{\psi}{\cal
D}\psi\exp i\{S+\int d^{4}x {\cal L}_{J}]\},
\end{eqnarray}
where the action $S$ is given in \eqref{(2.3)} and the source term is
${\cal L}_{J}=\bar{\psi}(x)\eta(x)+\bar{\eta}(x)\psi(x)$. The
$T^{\star}$-product
is quite different from the canonical $T$-product in the present
nonlocal theory, and in fact the canonical quantization is not defined
in the present theory. It is however important to note that the
$T^{\star}$-product can reproduce all the results of the  $T$-product,
if the $T$-product is well-defined, by means of the Bjorken--Johnson--Low
prescription~\cite{fujikawa}. In the present example, the presence of
the sine-function in the denominator of the correlation function
complicates this procedure, which is an indication of  the absence of
the canonical quantization of \eqref{(2.3)}. We also emphasize that the
analysis of the mass-splitting can be performed in terms of the exact
solution of the (modified) free Dirac equation \eqref{(2.4)}, which also
defines the propagator in the present path integral prescription.
 After all, Dirac discovered the antiparticle by solving his equation
exactly.
The propagator \eqref{(2.7)} is also an exact propagator for
\eqref{(2.3)} in the sense of the propagator theory of relativistic
quantum mechanics, and thus it could describe the particle and
antiparticle propagation if one understands the antiparticle with
negative energy propagating backward in time. However, if one attempts
to describe the particle and antiparticle propagation with definite
masses by pole approximation, for example, then the off-shell Lorentz
covariance of the propagator \eqref{(2.7)} is lost, as is discussed
later (see eq. \eqref{(2.17)}).

For the space-like $p$, the extra term with $\mu$ in the denominator of
the propagator \eqref{(2.7)} vanishes since $f_{+}(p)=f_{-}(p)$ for
$p=(0,\vec{p})$, as is shown in \eqref{(1.5)}. Thus the propagator has
poles only at the time-like momentum, and in this sense the present
Hermitian action \eqref{(2.3)} does not allow a tachyon.
By assuming a time-like $p$, we go to the frame where $\vec{p}=0$.
Then the eigenvalue equation is given by
\begin{eqnarray}\label{(2.9)}
p_{0}&=&\gamma_{0}\{m + i\mu [f_{+}(p_{0})-f_{-}(p_{0})]\},
\end{eqnarray}
namely,
\begin{eqnarray}\label{(2.10)}
p_{0}&=&\gamma_{0}\left[m - 4\pi \mu\int_{0}^{\infty}dz\frac{z^{2}\sin [
p_{0}\sqrt{z^{2}+l^{2}}]}{\sqrt{z^{2}+l^{2}}}\right],
\end{eqnarray}
where we used  the explicit formula in \eqref{(1.4)}.
The solution $p_{0}$ of this equation \eqref{(2.10)} determines the
possible mass eigenvalues.

This eigenvalue equation under $p_{0}\rightarrow -p_{0}$ becomes:
\begin{eqnarray}\label{(2.11)}
-p_{0}
&=&\gamma_{0}\left[m + 4\pi\mu\int_{0}^{\infty}dz\frac{z^{2}\sin [
p_{0}\sqrt{z^{2}+l^{2}}]}{\sqrt{z^{2}+l^{2}}}\right].
\end{eqnarray}
By sandwiching this equation by $\gamma_{5}$, which is regarded as CPT
operation, we have
\begin{eqnarray}\label{(2.12)}
-p_{0}
&=&\gamma_{0}\left[-m - 4\pi\mu\int_{0}^{\infty}dz\frac{z^{2}\sin [
p_{0}\sqrt{z^{2}+l^{2}}]}{\sqrt{z^{2}+l^{2}}}\right],
\end{eqnarray}
i.e.,
\begin{eqnarray}\label{(2.13)}
p_{0}
&=&\gamma_{0}\left[m + 4\pi\mu\int_{0}^{\infty}dz\frac{z^{2}\sin [
p_{0}\sqrt{z^{2}+l^{2}}]}{\sqrt{z^{2}+l^{2}}}\right],
\end{eqnarray}
which is not identical to the original equation in \eqref{(2.10)}.
In other words, if $p_{0}$ is the solution of the original equation,
$-p_{0}$ cannot be the solution of the original equation except for
$\mu=0$. The last term in the Lagrangian \eqref{(2.3)} with
C=CP=CPT=$-1$ splits the particle and antiparticle masses.

As a crude estimate of the mass splitting, one may assume $\mu\ll
m$ and solve these equations iteratively. If the particle mass for
\eqref{(2.10)} is chosen at
\begin{eqnarray}\label{(2.14)}
p_{0}\simeq m - 4\pi \mu\int_{0}^{\infty}dz\frac{z^{2}\sin [ m
\sqrt{z^{2}+l^{2}}]}{\sqrt{z^{2}+l^{2}}},
\end{eqnarray}
then the antiparticle mass for \eqref{(2.13)} is estimated at
\begin{eqnarray}\label{(2.15)}
p_{0}\simeq m + 4\pi \mu\int_{0}^{\infty}dz\frac{z^{2}\sin [ m
\sqrt{z^{2}+l^{2}}]}{\sqrt{z^{2}+l^{2}}}.
\end{eqnarray}

\subsection{ Canonical description}

Once one finds eigenvalues,  one may examine the behavior of the
off-shell propagator \eqref{(2.7)} around those pole positions
approximately and may apply the Bjorken--Johnson--Low (BJL) prescription
to reveal the canonical structure~\cite{fujikawa}. Then one finds an
operator description of those particle and antiparticle with different
masses, although the manifest invariance is lost. (This is somewhat
analogous to the electromagnetic field. The off-shell Maxwell equation
is manifestly Lorentz covariant but if one applies the physical Coulomb
gauge to define the photon, the manifest invariance is lost.)
We would like to explain the basic steps of this procedure.

If one denotes the particle mass by $m_{+}$ and antiparticle mass by
$m_{-}$, respectively,  we have {\em approximately} near the pole
positions of the propagator in \eqref{(2.7)}:
\begin{eqnarray}\label{(2.16)}
\int d^{4}x e^{ip(x-y)}\langle T^{\star}\psi(x)\bar{\psi}(y)\rangle&\simeq
&\frac{i}{\pslash-m_{+} +i\epsilon},  \ {\rm for} \ p_{0}>0,
\cr
&\simeq&\frac{i}{\pslash-m_{-} +i\epsilon}, \ {\rm for} \ p_{0}< 0.\cr
\end{eqnarray}
The first step of BJL prescription is to examine the large $p_{0}$
behavior of the right-hand side of the Fourier transform, which goes to
$0$ in the present case. In this case, we replace the
$T^{\star}$-product  by the canonical $T$-product.

From the point of view of the  field-product $\psi(x)\bar{\psi}(y)$,
$T$ is supposed to specify the product even for the precise coincident
time $x^{0}=y^{0}$, while $T^{\star}$ specifies the product only for
$x^{0}\neq y^{0}$ and examine the behavior of the product for
$x^{0}-y^{0}\rightarrow 0$ later. These two procedures agree with each
other for the theory where ordinary canonical quantization is well-defined,
but in general they do not agree with each other. The Schwinger term,
for example, is identified by this disagreement. If the short-time limit
is well-specified by $T$, the  Riemann-Lebesgue lemma in the Fourier
transform implies that the large frequency limit of the $T$-product
vanishes. This is the basis of the above replacement of $T^{\star}$ by
$T$~\cite{fujikawa}.

We thus have the relations, which are more specific than \eqref{(2.16)}
but still approximate, although we use the equality symbol:
\begin{eqnarray}\label{(2.17)}
\int d^{4}x e^{ip(x-y)}\langle T\psi_{+}(x)\bar{\psi}_{+}(y)\rangle=
\frac{i\Lambda_{+}(m_+)}{p^{2}-m^{2}_{+} +i\epsilon},\cr  \ \ \ {\rm for} \
\ p_{0}>0,
\nonumber\\
\int d^{4}x e^{ip(x-y)}\langle T\psi_{-}(x)\bar{\psi}_{-}(y)\rangle=
\frac{i\Lambda_{-}(m_-)}{p^{2}-m^{2}_{-} +i\epsilon}, \cr   \ \ \ {\rm for}
\ \ p_{0}< 0,
\end{eqnarray}
 where we have separated $\psi$ into positive $\psi_{+}(x)$ and negative
$\psi_{-}(x)$ frequency components. We used the positive energy and
negative energy projection operators constructed by the solutions of
\eqref{(2.6)}, where $\Lambda_{+}(m)+\Lambda_{-}(m)=\pslash +m$ for the
equal mass case. The final step of BJL prescription is to multiply
both-hand sides of the relations in \eqref{(2.17)} by $p_{0}$ and
consider the large $p_{0}$ limit. For example,
 \begin{eqnarray}\label{(2.18)}
&&p_{0}\int d^{4}x e^{ip(x-y)}\langle
T\psi_{+}(x)\bar{\psi}_{+}(y)\rangle\nonumber\\
&=&-i\int d^{4}x \frac{\partial}{\partial x^{0}}e^{ip(x-y)}\langle
T\psi_{+}(x)\bar{\psi}_{+}(y)\rangle\nonumber\\
&=&i\int d^{4}x e^{ip(x-y)}\frac{\partial}{\partial x^{0}}\langle
T\psi_{+}(x)\bar{\psi}_{+}(y)\rangle\nonumber\\
&=&i\int d^{4}x
e^{ip(x-y)}[\langle\delta(x^{0}-y^{0})\{\psi_{+}(x),\bar{\psi}_{+}(y)\}\rangle\nonumber\\
&+& \langle T\frac{\partial}{\partial
x^{0}}\psi_{+}(x)\bar{\psi}_{+}(y)\rangle]
=\frac{ip_{0}\Lambda_{+}(m_+)}{p^{2}-m^{2}_{+} +i\epsilon}.
\end{eqnarray}
We consider the limit $p_{0}\rightarrow \infty$ in this relation. The
$T$-product part $\langle T\frac{\partial}{\partial
x^{0}}\psi_{+}(x)\bar{\psi}_{+}(y)\rangle$ goes to $0$ in this limit
since it is always {\em defined} to
satisfy the Riemann-Lebesgue type condition, which specifies the
separation between the commutator part and the $T$-product part
uniquely. We thus conclude from the last two relations in \eqref{(2.18)}
\begin{eqnarray}\label{(2.19)}
&&\delta(x^{0}-y^{0})\{\psi_{+}(x),\psi^{\dagger}_{+}(y)\}=\frac{1}{2}\delta^{4}(x-y),\nonumber\\
&&\delta(x^{0}-y^{0})\{\psi_{-}(x),\psi^{\dagger}_{-}(y)\}=\frac{1}{2}\delta^{4}(x-y),
\end{eqnarray}
where the second relation follows from the second relation in
\eqref{(2.17)}.
Note that at extremely high energies, the mass difference does not
matter at least in the fixed mass approximation. By this way, we obtain
an approximate canonical description of the CPT-violated fermion with
mass splitting.  The basic approximation involved in the transition from
the manifestly Lorentz covariant off-shell propagator to the approximate
canonical description is the identification of the pole structure in
\eqref{(2.16)}, which is exact for the case of an identical
(momentum-independent) mass.

\section{Conclusion}

 We have presented a simple Lorentz invariant CPT violating Lagrangian
model in \eqref{(2.3)}, which produces the splitting of particle and
antiparticle masses. The simple Lagrangian model will provide a useful
theoretical laboratory when one investigates  Lorentz invariant CPT
violation effects.

Besides the fact that both CPT and Lorentz invariance are two most fundamental
symmetries in physics, whose violations have not been hitherto observed, the relation
between the two symmetries and their possible breaking are of considerable theoretical
and experimental interest. Recent MINOS neutrino experiments with their favoured
interpretation through a mass difference for muon neutrino  and antineutrino have
revived interest in CPT violation and its possible implication on Lorentz invariance
breaking \cite{murayama,altarelli,adamson}.

It is an interesting question whether the CPT violation
in our model could be a long distance effective description of some
modified structure of space-time at short distances, for example.


Our Lagrangian is nonlocal and the local gauge principle cannot be
directly applied to it. Nevertheless, its novelty is the specific realization of a $CPT$-odd and Lorentz invariant model, showing a mass splitting of fermions and, in general, of any particle and its antiparticle.

\section*{Acknowledgements}

One of the authors (KF) is grateful to the Department of Physics, University
of Helsinki, for the hospitality during the summer of 2011,
when the present work was initiated.

The support of the Academy
of Finland under the Projects No. 136539 and 140886, and the support for this
research from the Magnus Ehrnrooth Foundation, Finland, are
gratefully acknowledged.



\end{document}